\documentclass[aps,prl,twocolumn,10pt,groupedaddress]{revtex4-1}
\usepackage{amssymb,amsfonts,amsmath}
\usepackage[dvips]{graphicx}
\usepackage{epsfig}

\begin{document}

\title{The Panic Model: Flocking with Minimal Cooperativity}

\author{Kevin R. Pilkiewicz}\affiliation{University of Colorado Boulder}
\author{Joel D. Eaves}\affiliation{University of Colorado Boulder}

\date{\today}

\begin{abstract}
We present a 2D lattice model of self-propelled spins that can only change direction upon collision with another spin. We show that even with ballistic motion and minimal cooperativity, these spins display robust flocking behavior at nearly all densities, forming long bands of stripes. The structural transition in this system can be characterized by an order parameter, and we demonstrate that if this parameter is studied as a dynamical variable rather than a steady-state observable, we can extract a detailed picture of how the flocking mechanism varies with density.
\end{abstract}

\maketitle

Flocking is ubiquitous in nature, exhibited by a diverse host of organisms ranging from bacteria in the human body to wildebeests on the Serengeti.  While ``strength in numbers'' can be seen as an evolutionary impetus for flocking behavior in animals, the spectrum of flocking fauna run the gamut of intelligence and biological complexity, and they have disparate capacities to sense and react to their external environments. This suggests that flocking must have its origin in simple mechanical tendencies that do not depend upon some sort of survival instinct. The observation that similar behaviors occur in a variety of nonliving systems such as driven colloids\cite{Reichhardts,Surfers} and granular media\cite{VijayRama,GranRods,GranDisks}, active polar gels\cite{Groswassers,Prost1,Toyota1}, and robot swarms\cite{YoChong,AliSahin} further supports this notion. 

Many theoretical models have been developed to study flocking, including self-propelled particle models in continuous space\cite{VicsekModel,OhOhta,Hagan1,FlyingXY,Moshers} and on a lattice\cite{RaymondEvans,TrafficJams} as well as hydrodynamic models\cite{TonerTu,Prost2,Gregoire1,Hagan2}. Most of these models agree that the principal driving force for flocking is a local, isotropic interaction between flockers that favors alignment of their velocities. This putative isotropy is problematic, however, in that it allows members of a flock to react to the motion of the members behind them just as they do to those in front of them. This effectively precludes any sort of ``follow the leader'' behavior, frequently observed in the flocking of animals like ants and sheep. Since most organisms cannot sense their environment equally well in all directions, robust flocking cannot depend on such a high level of cooperativity.

In this letter, we demonstrate that flocking can occur with high fidelity, even with minimal cooperativity between agents. We present a two dimensional lattice model where the interactions between flockers are markedly anisotropic and extremely short-ranged. It consists of self-propelled spins that move ballistically, exclude volume, and can only ``see'' directly in front of them, effectively minimizing the cooperativity present in the dynamics. The behavior of these spins is reminiscent of that of panicked animals, so we call this model the Panic Model. We will show that this model exhibits a spontaneous flocking transition for a broad range of densities, and that this transition can be couched in the language of thermodynamic phase transitions by introduction of an appropriate order parameter. Because flocking is a global phenomenon spawned by local interactions, large, system spanning structures must coalesce out of a smaller scale coarsening; and we will show that this same order parameter can be studied as a dynamical variable to elucidate how the local coarsening of the system proceeds to the global steady-state as a function of density.

Our model consists of an $L\times L$, periodic square lattice with a fixed fraction $\rho$ of its sites occupied by spins that can take one of four orientations--up, down, left, or right--such that no two spins occupy the same site. Each spin can only move in the direction of its orientation, and it will move ballistically in that direction, one lattice spacing per time step, until another spin blocks its path. When this happens, the spin will attempt to change its orientation to match that of the obstructing spin, though it will fail to do so with a specified probability.

The dynamics of the Panic Model proceed according to the following algorithm: First, the system is initialized in a random configuration with density $\rho$. At the beginning of each time step, a spin is chosen at random to attempt a move of one lattice spacing in the direction of its orientation. This move is always accepted unless it would place two spins on the same site. In that case the move fails and the selected spin instead changes its orientation to that of the spin obstructing its movement plus a clockwise angle $\theta$ that can be either 0, $\pm\pi/2$, or $\pi$. The probability that this angle takes a given value is determined by a distribution $P_{\theta}$ that we parametrize by an error parameter $0\leq\epsilon<\infty$ that is fixed for each simulation and acts as an ersatz temperature.
\begin{align}
P_{0}(\epsilon)&=\frac{1}{4}+\frac{3/4}{1+(1/10)\epsilon^2},\,\,\,
P_{\pm\pi/2}(\epsilon)=\frac{1}{4}-\frac{1/4}{1+(3/25)\epsilon^2} \nonumber \\
P_{\pi}(\epsilon)&=1-P_{0}(\epsilon)-2P_{\pm\pi/2}(\epsilon)
\end{align} 
Another spin is then chosen at random from the remaining spins, and the previous steps are repeated. Once each spin has had a chance to act, the clock is advanced by one time step. Note that at $\epsilon=0$, spins always align with the spins that block their movement, but as $\epsilon$ is increased, they change their orientation more and more randomly (though for finite $\epsilon$, a spin is always more likely to make a smaller error than a larger one).

We simulated the Panic Model for $L=100$ over a broad range of $(\rho,\epsilon)$ parameter space. At sufficiently large $\epsilon$, the system remains disordered for all time; but for all densities exceeding roughly $\rho=0.07$, there was a density dependent, threshold value of $\epsilon$ below which the system was observed to self-organize into bands of either vertical or horizontal stripes, spontaneously breaking the discrete rotational symmetry of the lattice. Importantly, the dynamical rules of the Panic Model destabilize a host of ordered steady-state phases observed in other models of flocking, such as asters, waves, and gliders\cite{FlyingXY,TrafficJams}. These mutually aligned striped domains are thus the only stable ordered phase observed at steady-state.

\begin{figure}[ht!]
\includegraphics[width=8.5cm,height=6.5cm,keepaspectratio=true]{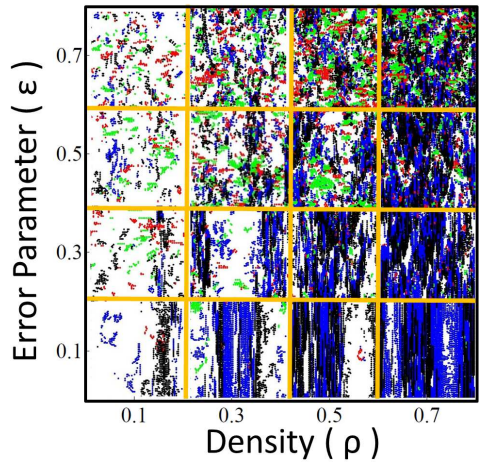}
\caption{(Color online) Steady-state snapshots from Panic Model simulations at various points in $(\rho,\epsilon)$ parameter space. In each snapshot, up, down, left, and right spins are represented by black, blue, red, and green dots, respectively.}
\end{figure}

Snapshots of typical steady-state configurations of the system for a range of $\rho$ and $\epsilon$ can be seen in Fig. 1. In these images, the spins are represented as colored points, with black, blue, red, and green points representing up, down, left, and right spins, respectively. To make the divide between ordered and disordered states more pronounced, we depicted only vertically striped steady-states in this figure, where all spins are oriented either up or down. It is evident from the figure that the threshold value of $\epsilon$ is a monotonically increasing function of the density.

Just like a thermodynamic phase transition, the Panic Model structural transition can be characterized by an order parameter that is zero in the disordered state and unity in a perfectly striped state.
\begin{align}
\sigma(\rho,\epsilon,t) &= \langle\left\vert\vphantom{\frac{1}{1}}\left(\chi_{\uparrow}(\rho,\epsilon,t)+\chi_{\downarrow}(\rho,\epsilon,t)\right)\right. \nonumber \\
&\left.-\left(\chi_{\leftarrow}(\rho,\epsilon,t)+\chi_{\rightarrow}(\rho,\epsilon,t)\right)\vphantom{\frac{1}{1}}\right\vert\rangle
\end{align}
In this expression, each $\chi$ is the fraction of spins with the indicated orientation at time $t$ for a given configuration of the system, and the angular brackets denote an average over all configurations of the system attainable at density $\rho$ and error parameter $\epsilon$ after a time $t$, assuming all configurations are equally likely at time $t=0$. In principle, this is a very difficult average to compute, but the distribution of $\sigma$ is sufficiently peaked about its mean that it is adequate in practice to average over only a couple thousand configurations.

\begin{figure}[ht!]
\includegraphics[width=8.5cm,height=4.25cm,keepaspectratio=true]{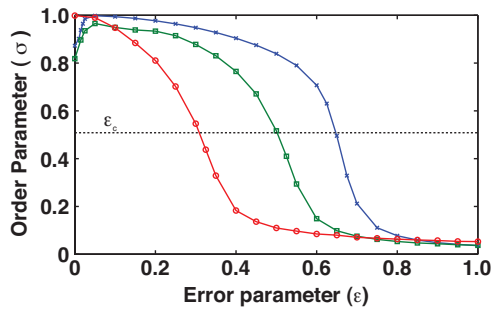}
\caption{(Color online) The order parameter, $\sigma$, at steady-state ($t=$20,000 time steps) plotted as a function of the error parameter, $\epsilon$, for densities of 0.15 (red O's), 0.45 (green $\Box$'s), and 0.85 (blue X's). We define the critical $\epsilon$ to be where the dashed line, $\sigma=0.5$, crosses each curve.}
\end{figure}

\begin{figure}[ht!]
\includegraphics[width=8.5cm,height=4.25cm,keepaspectratio=true]{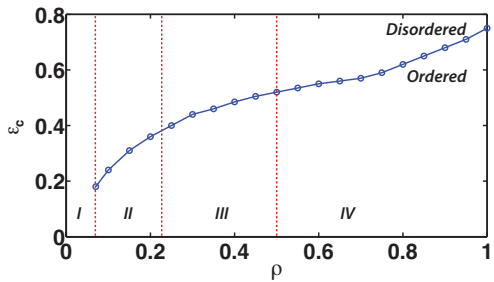}
\caption{(Color online) The phase diagram for the Panic Model. The critical error parameter $\epsilon_c$ plotted as a function of density separates the disordered and ordered phases. Dashed vertical lines at $\rho=0.07$, $0.225$, and $0.50$ partition the structural transition into distinct dynamical regimes, labeled by Roman numerals.}
\end{figure}

\begin{figure*}[ht!]
\includegraphics[width=17cm,height=17cm,keepaspectratio=true]{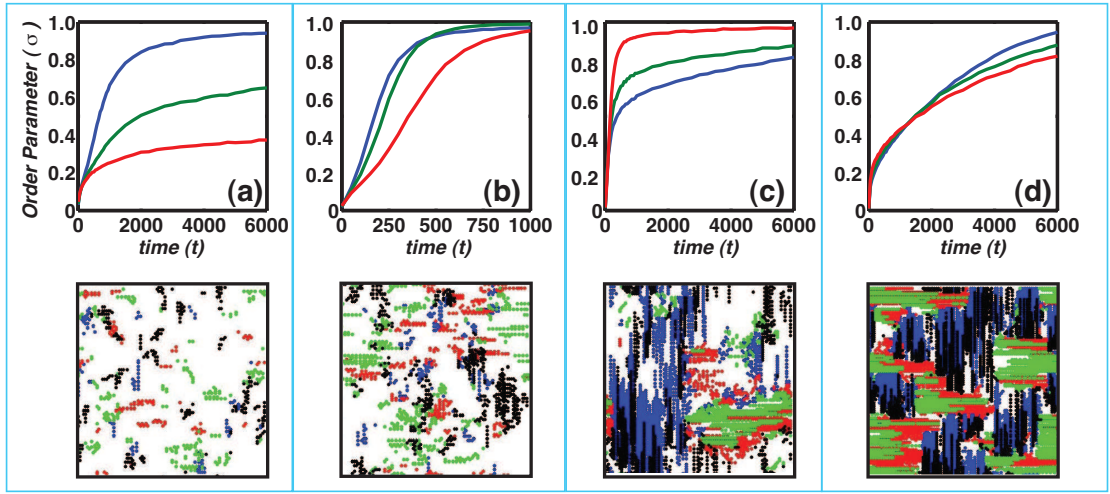}
\caption{(Color online) The order parameter, $\sigma$, plotted as a function of time (a) in regime I, for densities of 0.03 (red), 0.05 (green), and 0.07 (blue); (b) in regime II, for densities of 0.10 (red), 0.15 (green), and 0.20 (blue); (c) in regime III, for densities of 0.25 (red), 0.35 (green), 0.40 (blue); (d) and in regime IV, for densities of 0.60 (red), 0.75 (green), and 0.95 (blue). The lower half of each panel depicts a typical configuration of the system after 100 time steps in the corresponding dynamical regime (spins are color coded the same as in Fig. 1).}
\end{figure*} 

Figure 2 is a plot of the order parameter, $\sigma$, at steady-state as a function of $\epsilon$ for three different densities. For low densities, $\sigma$ behaves just like the order parameter of a finite sized thermodynamic system as temperature lowers across a phase boundary. The order parameter saturates near unity for small $\epsilon$, approaches zero for large $\epsilon$, and changes very sharply over a narrow range of $\epsilon$ somewhere in between. For larger densities, the behavior is the same, except that for very small $\epsilon$ (roughly $\epsilon<0.05$), the order parameter falls sharply away from saturation. This anomaly comes from the contribution of frustrated configurations in which a single band of stripes that is orthogonal to the others persists at steady-state either because it is system spanning itself or because it is trapped between two system spanning bands perpendicular to it. These configurations cannot break down, even at very long times, when the error parameter is too small. At lower densities, gaps in the stripes preclude this kind of frustration from occurring.

Since our system is finite in size, the structural transition to a striped steady-state actually occurs over a small range of $\epsilon$ rather than below a single critical value, but to a good approximation we can define a critical error parameter, $\epsilon_c$, as the value of $\epsilon$ for which $\sigma=0.50$ at steady-state. Plotting $\epsilon_c$ versus the density yields a phase diagram for the Panic Model (Figure 3) consisting of a single phase curve above which the system has a disordered steady-state and below which it has an ordered, striped steady-state. The phase curve terminates at $\rho=0.07$ because below this density the order parameter does not saturate towards unity.

We probe how the dynamics of this structural transition vary over such a broad range of densities by studying the time evolution of $\sigma$ as the system evolves from a random configuration to a striped steady-state. Four qualitatively distinct types of behavior are observed, allowing us to divide the phase diagram into four dynamical regimes (demarcated by vertical dashed lines in Fig. 3 and labeled by Roman numerals). The characteristic time evolution of the order parameter in each of these regimes is shown in Fig. 4.

Figure 4(a) shows the order parameter at zero $\epsilon$ plotted as a function of time for three densities in regime I of the phase diagram. The order parameter grows steadily for roughly the first thousand time steps, corresponding to an initial coarsening of the system into small unidirectional clusters, as shown in the lower panel for one particular simulation at $\rho=0.07$ and $t=100$ time steps. As these clusters begin to grow in size, the density exhibits large fluctuations in space, making it possible to end up with orthogonally moving clusters that can never collide, preventing the transition to a fully striped steady-state and causing $\sigma$ to taper off to a value less than unity.

In Figure 4(b), the time evolution of the order parameter in region II is depicted, again at $\epsilon=0$ for three densities. Due to the higher density, the clusters that initially form in this regime are larger in size, as can be seen from the lower panel, which depicts a typical configuration at $\rho=0.15$ and $t=100$ time steps. Larger clusters can collide more facilely and will swiftly coalesce into stripes, as reflected by the rapid saturation of the order parameter. Though there is a monotonic increase in the growth rate of $\sigma$ with density, its ultimate saturation begins to occur more slowly as regime III is approached.

The analogous plot for regime III is shown in Fig. 4(c), this time for $\epsilon=0.05$, where $\sigma$ has its maximum steady-state value at these densities (see Fig. 2). In this regime, the order parameter rises very sharply over the first roughly hundred time steps, which corresponds to the system rapidly coalescing into large, competing domains of horizontally and vertically oriented spins, as seen in the lower panel for a simulation at $\rho=0.40$ and $t=100$ time steps. The large, often system spanning size of these domains makes them difficult to break down, frustrating the completion of the structural transition and resulting in a drastic slowdown in the growth of $\sigma$ at longer times. As density is increased, the transition between the fast and slow growth regimes occurs at smaller values of $\sigma$. The rate of growth in the slow regime does increase with density, but lower density systems still saturate faster overall.

The growth of the order parameter in the highest density regime is plotted in Fig. 4(d), once again at $\epsilon=0.05$ for three different densities. A snapshot of the system at $\rho=0.75$ and $t=100$ time steps (see the lower panel) shows that the domains that initially emerge in this regime are smaller than those of the previous regime. Smaller domains are easier to break down, so the overall saturation is faster in this regime, though still much slower than in regime II. In this regime the behavior of $\sigma$ is well fit by a simple power law with exponent less than one that increases monotonically with density. This means that saturation occurs more quickly for denser systems, a result of the domains having more frequent collisions at their boundaries due to being packed more tightly together. This effect is also present in regime III, but there it is offset by a steady increase in the size of the domains with higher density.

The feature of having a single structural transition with varying dynamical character is reminiscent of thermodynamic phase transitions in other systems. For example, the Blume-Capel model\cite{Blume1,Blume2}, an Ising model variant, exhibits a single phase transition to a ferromagnetic state, but depending on where the phase curve is crossed, the transition can be either first or second order. By analogy, the Panic Model structural transition occurs through a nucleation process in regime II that is reminiscent of the sorts of activated processes present in freezing liquids (a first order phase transition), whereas in regime III and IV the transition occurs through the formation of competing domains, not unlike the second order ferromagnetic transition of the standard Ising model. The point on the phase curve that separates regime II from regime III ($\rho\approx 0.225$) is thus something akin to a tricritical point connecting a line of first order-like transitions to one of second order-like transitions.

The Panic Model illustrates that self-propelled particles with a minimal amount of cooperativity can still exhibit robust flocking behavior. The sort of collision-based flocking that this model describes depends heavily upon excluded volume interactions, and as such it is not surprising that the dynamics of this flocking are highly sensitive to the density of the system. It is likely that this sensitivity is also present in more cooperative flocking models with excluded volume interactions, but in those models it is occluded by the larger variety of steady-state structures that can emerge at different densities. 

It is tempting to focus on the steady-state structural features of flocking models because they can be related to better understood equilibrium phenomena like phase transitions, but to truly understand far from equilibrium phenomena like flocking, it will be necessary to develop new tools for studying and describing these systems at times other than zero and infinity. The regimented dynamical behavior of the order parameter in our model suggests that such parameters, traditionally studied only as steady-state observables, may be a useful tool for studying the dynamics of coarsening and self-assembly in other, more complex systems as well.

\end{document}